\documentclass[10pt,conference]{IEEEtran}

\usepackage[T1]{fontenc}
\usepackage{cite}
\usepackage[final,bookmarks=false]{hyperref}
\usepackage[USenglish]{babel}
\usepackage[cmex10]{amsmath}
\usepackage{amsthm}
\usepackage{amsfonts}
\usepackage{amssymb}
\usepackage{mathtools}
\usepackage[ascii]{inputenc}
\usepackage{aliascnt}
\usepackage[all]{hypcap}
\usepackage{algpseudocode}
\usepackage{array}
\usepackage{url}

\newtheorem*{thm-plain}{Theorem}
\newtheorem{thm}{Theorem}[section]
\newcommand{\newjointcountertheorem}[3]{\newaliascnt{#1}{#2}\newtheorem{#1}[#1]{#3}\aliascntresetthe{#1}}
\newjointcountertheorem{satz}{thm}{Satz}
\newjointcountertheorem{lem}{thm}{Lemma}
\newjointcountertheorem{cor}{thm}{Corollary}
\newjointcountertheorem{prp}{thm}{Proposition}
\newjointcountertheorem{cnj}{thm}{Conjecture}
\newjointcountertheorem{que}{thm}{Question}
\newjointcountertheorem{prb}{thm}{Problem}
\newjointcountertheorem{fct}{thm}{Fact}
\newjointcountertheorem{obs}{thm}{Observation}
\newjointcountertheorem{alg}{thm}{Algorithm}
\newjointcountertheorem{asm}{thm}{Assumption}
\theoremstyle{definition}
\newjointcountertheorem{dfn}{thm}{Definition}
\newjointcountertheorem{ntn}{thm}{Notation}
\theoremstyle{remark}
\newjointcountertheorem{rem}{thm}{Remark}
\newjointcountertheorem{nte}{thm}{Note}
\newjointcountertheorem{exl}{thm}{Example}

\def\Snospace~{\S{}}

\DeclareMathOperator{\ch}{ch}

\DeclareMathOperator{\Sym}{Sym}

\DeclareMathOperator{\DuHe}{DH}
\DeclareMathOperator{\aff}{aff}

\DeclareMathOperator{\U}{U}

\DeclareMathOperator{\Lie}{Lie}

\DeclareMathOperator{\poly}{poly}
\DeclareMathOperator{\sesi}{ss}

\newcommand{\PP}{\mathbb P}  
\newcommand{\CC}{\mathbb C}
\newcommand{\RR}{\mathbb R}

\newcommand{\ZZ}{\mathbb Z}

\newcommand{\Id}{\mathbf 1}
\renewcommand{\P}{\mathbf P}  
\newcommand{\NP}{\mathbf{NP}}
\newcommand{\SharpP}{\#\mathbf P}
\newcommand{\GapP}{\mathbf{GapP}}

\newcommand{\abs}[1]{\lvert#1\rvert}
\newcommand{\restrict}[2]{\left.#1\vphantom{\big|}\right|_{#2}}
\renewcommand{\matrix}[1]{\left(\begin{smallmatrix}#1\end{smallmatrix}\right)}
\newcommand{\Matrix}[1]{\begin{pmatrix}#1\end{pmatrix}}
\newcommand{\su}{\mathfrak{su}}

\numberwithin{equation}{section}
\begin{document}

\title{Computing Multiplicities \\ of Lie Group Representations}


\author{\IEEEauthorblockN{Matthias Christandl}
\IEEEauthorblockA{Department of Physics \\
ETH Z\"urich\\
Z\"urich, Switzerland\\
Email: christandl@phys.ethz.ch}
\and
\IEEEauthorblockN{Brent Doran}
\IEEEauthorblockA{Department of Mathematics \\
ETH Z\"urich\\
Z\"urich, Switzerland\\
Email: brent.doran@math.ethz.ch}
\and
\IEEEauthorblockN{Michael Walter}
\IEEEauthorblockA{Department of Physics \\
ETH Z\"urich\\
Z\"urich, Switzerland\\
Email: mwalter@itp.phys.ethz.ch}
}


\maketitle

\begin{abstract}
For fixed compact connected Lie groups H~$\subseteq$~G, we provide a polynomial time algorithm to compute the multiplicity of a given irreducible representation of H in the restriction of an irreducible representation of G. Our algorithm is based on a finite difference formula which makes the multiplicities amenable to Barvinok's algorithm for counting integral points in polytopes.

The Kronecker coefficients of the symmetric group, which can be seen to be a special case of such multiplicities, play an important role in the geometric complexity theory approach to the P vs.\ NP problem. Whereas their computation is known to be \#P-hard for Young diagrams with an arbitrary number of rows, our algorithm computes them in polynomial time if the number of rows is bounded.
We complement our work by showing that information on the asymptotic growth rates of multiplicities in the coordinate rings of orbit closures does not directly lead to new complexity-theoretic obstructions beyond what can be obtained from the moment polytopes of the orbit closures. Non-asymptotic information on the multiplicities, such as provided by our algorithm, may therefore be essential in order to find obstructions in geometric complexity theory.
\end{abstract}

\IEEEpeerreviewmaketitle

\section{Introduction}
\label{section:introduction}

The decomposition of Lie group representations into irreducible sub-representations is a fundamental problem of mathematics with a variety of applications to the sciences. 
In atomic and molecular physics (Clebsch--Gordan series), as well as in high-energy physics, this problem has been studied extensively \cite{weyl50,wigner59,wigner73}, perhaps most famously in Ne'eman and Gell-Mann's eight-fold way of elementary particles \cite{neeman,gellmann2,gellmann}.
In pure mathematics, the combinatorial resolution of the problem of decomposing tensor products of irreducible representations of the unitary group by Knutson and Tao has been a recent highlight with a long history of research \cite{fulton00,knutsontao99}. 
More recently, the theories of quantum information \cite{keylwerner01,christandlmitchison06,klyachko06}, computation and complexity \cite{baconchuangharrow07}, as well as the geometric complexity theory approach to the $\P$ vs.\ $\NP$ problem \cite{mulmuleysohoni01,mulmuleysohoni08,burgisserlandsbergmaniveletal11} have brought the representation theory of Lie groups to the attention of the computer science community.

In this paper, we study the problem of computing multiplicities of Lie group representations:

\begin{prb}[Subgroup Restriction Problem]
  \label{main problem}
  Let $f \colon H \rightarrow G$ be a homomorphism between compact connected Lie groups $H$ and $G$.
  The \emph{subgroup restriction problem for $f$} is to determine the multiplicity $m^\lambda_\mu$ of the irreducible $H$-representation $V_{H,\mu}$ in the irreducible $G$-representation $V_{G,\lambda}$ when given as input the highest weights $\mu$ and $\lambda$ (specified as bitstrings containing their coordinates with respect to fixed bases of fundamental weights, see \autoref{section:algorithm}).
\end{prb}

The name \emph{subgroup restriction problem} comes from the archetypical case where the map $f$ is induced by the inclusion of a subgroup $H \subseteq G$. \autoref{main problem} is also known as the \emph{branching problem}.
The main result of this paper is a polynomial-time algorithm for \autoref{main problem}:

\begin{thm}
  \label{A}
  For any homomorphism $f \colon H \rightarrow G$ between compact connected Lie groups $H$ and $G$, there is a polynomial-time algorithm for the subgroup restriction problem for $f$.
\end{thm}

Indeed, we describe a concrete algorithm (\autoref{main algorithm}). In particular, for any fixed $\lambda$ and $\mu$ the stretching function $k \mapsto m^{k \lambda}_{k \mu}$ can be evaluated in polynomial time.

\begin{cor}
  \label{Aprime}
  For any homomorphism $f \colon H \rightarrow G$ between compact connected Lie groups $H$ and $G$, positivity of the coefficients $m^\lambda_\mu$ can be decided in polynomial time.
\end{cor}

Mulmuley conjectures that deciding positivity of the multiplicities $m^\lambda_\mu$ is possible in polynomial time if the group homomorphism $f$ is also part of the input \cite{mulmuley07}. \autoref{Aprime} can be regarded as supporting evidence that this conjecture might in fact be true for general $f$ (note that for specific families of homomorphisms, such as those corresponding to the Littlewood--Richardson coefficients, positivity can be decided in polynomial time \cite{knutsontao99,mulmuleysohoni05}). However, any approach to deciding positivity that proceeds by computing the actual multiplicities is of course expected to fail, since the latter problem is well-known to be $\SharpP$-hard \cite{narayanan06,burgisserikenmeyer08}.

We establish \autoref{A} by deriving a novel formula for the multiplicities $m^\lambda_\mu$ (\autoref {main theorem}), which is obtained in three steps: First, we restrict from the group $G$ to its maximal torus $T_G$; the corresponding weight multiplicities can be computed efficiently by using the classical Kostant multiplicity formula \cite{kostant59,cochet05} or in fact by evaluating a single vector partition function \cite{billeyguilleminrassart04,bliem08,bliem10} (\autoref{section:preliminaries}). Second, we restrict all weights to a maximal torus $T_H$ of $H$. Third, we recover the multiplicity of an irreducible $H$-representation by using a finite-difference formula (\autoref{steinberg lemma}).
By carefully combining the first two steps, \autoref{main problem} can be reduced to counting integral points in certain rational convex polytopes of bounded dimension, which can be done efficiently by using Barvinok's algorithm \cite{barvinok94,dyerkannan97,barvinokpommersheim99} (see also \cite{dyer91,cook92,welledabaldonibeckcochetetal06}).

The multiplicity formula itself has intrinsic interest beyond its application to algorithmics. One insight that is immediate from our result is the piecewise quasi-polynomial nature of the multiplicities $m^\lambda_\mu$ (\autoref{abstract cor}).

\bigskip

Let us now turn to the computation of the \emph{Kronecker coefficients} $g_{\lambda,\mu,\nu}$, which arise in the decomposition of tensor products of irreducible representations of the symmetric group $S_k$ \cite{fulton97}:
\begin{equation*}
  [\lambda] \otimes [\mu] = \bigoplus_\nu g_{\lambda,\mu,\nu} \, [\nu],
\end{equation*}
where we denote by $[\lambda]$ the irreducible representation of $S_k$ labeled by the Young diagram $\lambda$ with $k$ boxes (\autoref{section:kronecker}).
Kronecker coefficients are notoriously difficult to study, and finding an appropriately strong combinatorial interpretation is one of the outstanding problems of classical representation theory. They appear naturally in geometric complexity theory, where their efficient computation has been subject to various conjectures \cite{mulmuley07}, as well as in quantum information theory in the context of the marginal problem and coding theory \cite{christandlmitchison06,daftuarhayden04,klyachko04,klyachko06,christandlharrowmitchison07,harrow05}.

Using Schur--Weyl duality, the Kronecker coefficients for Young diagrams with a bounded number of rows can be equivalently characterized in terms of a single subgroup restriction problem for compact connected Lie groups (\autoref{section:kronecker}). Therefore, by \autoref{A} they can also be computed efficiently:

\begin{cor}
  \label{B}
  For any fixed $d \in \ZZ_{> 0}$, there exists a polynomial-time algorithm for computing the Kronecker coefficient $g_{\lambda,\mu,\nu}$ given as input Young diagrams $\lambda$, $\mu$ and $\nu$ with at most $d$ rows. That is, the algorithm runs in $O(\poly(\log k))$ where $k$ is the number of boxes of the Young diagrams.
\end{cor}

\begin{cor}
  Positivity of Kronecker coefficients for Young diagrams with a bounded number of rows can be decided in polynomial time.
\end{cor}

By specializing our technique, we get a clean closed-form expression for the Kronecker coefficients (\autoref{optimized kronecker}), which not only nicely illustrates its effectiveness, but also implies piecewise quasi-polynomiality for bounded height (a feature that has only been noticed in a special case \cite{briandorellanarosas09}). Moreover, it is immediate from our formula that the problem of computing Kronecker coefficients with unbounded height is in $\GapP$, as first proved in \cite{burgisserikenmeyer08}.

Similar conclusions can be drawn for the plethysm coefficients, which can also be formulated in terms of subgroup restriction problems \cite{fultonharris91}. Like the Kronecker coefficients, they play a fundamental role in geometric complexity theory \cite{burgisserlandsbergmaniveletal11,burgisserchristandlikenmeyer11b} and quantum information theory \cite{klyachko06,christandlschuchwinter10}.

\bigskip

In practice, our algorithms appear to be rather fast as long as the rank of the Lie group $G$ is not too large. In the case of Kronecker coefficients for Young diagrams with two rows, we can easily go up to $k=10^8$ boxes using commodity hardware. In contrast, all other software packages known to the authors cannot go beyond only a moderate number of boxes ($k=10^2$ on the same hardware as used above). Moreover, by distributing the computation of weight multiplicities onto several processors, we have been able to compute Kronecker coefficients for Young diagrams with three rows and $k=10^5$ boxes in a couple of minutes.\footnote{A preliminary implementation of the algorithm is available upon request from the authors.} We hope that our algorithm will provide a useful tool in experimental mathematics, theoretical physics, and geometric complexity theory.

\bigskip

Our final result concerns the asymptotics of multiplicities in the general algebro-geometric setup of the geometric complexity theory approach to proving the $\mathbf{VP} \neq \mathbf{VNP}$ conjecture, an algebraic version of the $\P \neq \NP$ conjecture.
Recall that, in a nutshell, this approach amounts to showing that for certain pairs of projective subvarieties $X$ and $Y$ one is not contained in the other; this would then imply complexity-theoretic lower bounds. Both the permanent vs.\ determinant problem, which is equivalent to the $\mathbf{VP}$ vs.\ $\mathbf{VNP}$ problem \cite{valiant79}, as well as the complexity of matrix multiplication \cite{strassen69} can be formulated in this framework \cite{mulmuleysohoni01,mulmuleysohoni08,burgisserikenmeyer11,burgisserlandsbergmaniveletal11}.
More concretely, let us denote by $m_{H,X,k}(\mu)$ the multiplicity of the dual of an irreducible $H$-representation $V_{H,\mu}$ in the $k$-th graded part of the coordinate ring of $X$, and similarly for $Y$ (cf.\ \autoref{section:asymptotics} for precise definitions). Then,
\begin{equation}
  \label{mult crit}
  X \subseteq Y
  \;\Rightarrow\;
  m_{H,X,k}(\mu) \leq m_{H,Y,k}(\mu)
\end{equation}
for all $\mu$ and $k \geq 0$. Therefore, the existence of $\mu$ and $k$ such that $m_{H,X,k}(\mu) > m_{H,Y,k}(\mu)$ proves that $X \not\subseteq Y$; such a pair $(\mu,k)$ is called an \emph{obstruction} \cite{mulmuleysohoni08}. One can relax this implication further and instead compare the support of the multiplicity functions,
\begin{equation*}
  X \subseteq Y
  \;\Rightarrow\;
  \left( m_{H,X,k}(\mu) \neq 0 \Rightarrow m_{H,Y,k}(\mu) \neq 0 \right).
\end{equation*}
Since computing multiplicities in general coordinate rings is a difficult problem, it is natural to instead study their asymptotic behavior. Following an idea of Strassen \cite{strassen}, it has been proposed in \cite{burgisserikenmeyer11} to consider the \emph{moment polytope},
\begin{equation*}
  \Delta_X := \overline {\bigcup_{k=1}^\infty \left\{ \frac \mu k : m_{H,X,k}(\mu) \neq 0 \right\}},
\end{equation*}
which is a compact convex polytope that represents the asymptotic support of the stretching function. Moment polytopes do have a geometric interpretation, which should facilitate their computation \cite{brion87}. Clearly,
\begin{equation}
  \label{mo po crit}
  X \subseteq Y
  \;\Rightarrow\;
  \Delta_X \subseteq \Delta_Y.
\end{equation}
However, preliminary results suggest that the right-hand side moment polytope $\Delta_Y$ might be trivially large in the cases of interest \cite{burgisserchristandlikenmeyer11,burgisserikenmeyer11,kumar11,burgisserlandsbergmaniveletal11}, and therefore insufficient for finding complexity-theoretic obstructions.

It has therefore recently been suggested to study the \emph{asymptotic growth} of multiplicities (e.g., \cite[\S 2.2]{grochowrusek12}). The natural object is the \emph{Duistermaat--Heckman measure}, which is defined as the weak limit
\begin{equation}
  \label{definition duistermaat-heckman}
  \DuHe_X := \lim_{k \rightarrow \infty}  \frac 1 {k^{d_X}}  \sum_{\mu \in \Lambda^*_{H,+}} m_{H,X,k}(\mu) \, \delta_{\mu/k},
\end{equation}
where $d_X \in \ZZ_{\geq 0}$ is the appropriate exponent such that $\DuHe_X$ is a non-zero finite measure \cite{okounkov96}. The Duistermaat--Heckman measure has a continuous density function $f_X$ with respect to Lebesgue measure on the moment polytope; it is supported on the entire moment polytope (both statements follow from the main result of \cite{okounkov96}).
For well-behaved varieties, Duistermaat--Heckman measures have a geometric interpretation \cite{heckman82,guilleminsternberg82b,sjamaar95,meinrenken96,meinrenkensjamaar99,vergne98,teleman00}, which makes their computation potentially much more tractable \cite{guilleminlermansternberg88,guilleminlermansternberg96,christandldorankousidiswalter12} (this connection is however less clear in the singular cases relevant to geometric complexity theory). In this context, our main technical result is the following (see \autoref{section:asymptotics} for the proof):

\begin{thm}
  \label{C}
  The exponent $d_X$ is equal to $\dim X - R_X$, where $R_X$ is the number of positive roots of $H$ that are not orthogonal to all points of the moment polytope $\Delta_X$.
\end{thm}

The significance of \autoref{C} is that the order of growth of the ``smoothed'' multiplicities, as captured by the Duistermaat--Heckman measures, does only depend on the dimension of the orbit closures and on their moment polytopes.

Now suppose that we are in the situation that $X$ and $Y$ cannot be separated by using moment polytopes, i.e., $\Delta_X \subseteq \Delta_Y$. For the orbit closures $X$ and $Y$ that one tries to separate in geometric complexity theory, one can show that $\dim X < \dim Y$ \cite{burgisserlandsbergmaniveletal11,burgisserikenmeyer11}. Then, $X \subseteq Y$ would imply that $d_X < d_Y$ (\autoref{dimension lemma}). But this means that we \emph{cannot} deduce from \eqref{mult crit} and \eqref{definition duistermaat-heckman} a criterion of the form
\begin{equation*}
  X \subseteq Y
  \;\Rightarrow\;
  f_X(\mu) \leq f_Y(\mu)
  \qquad
  (\forall \mu),
\end{equation*}
since in order to take the weak limit we need to divide by different powers of $k$.
Therefore, Duistermaat--Heckman measures do not directly give rise to new obstructions, indicating that a more refined understanding of the behavior of multiplicities in coordinate rings might be required.

\section{Preliminaries}
\label{section:preliminaries}


In this paper we will use basic notions of the theory of compact Lie groups \cite{fultonharris91,cartersegalmacdonald95,kirillov08,knapp02}.
Let $G$ be a compact connected Lie group with Lie algebra $\mathfrak g$.
We fix a maximal torus $T_G \subseteq G$ and denote by $\mathfrak t_G$ its Lie algebra, the corresponding Cartan subalgebra. We write $\Lambda_G = \ker \restrict \exp {\mathfrak t_G}$ for the integral lattice and $\Lambda^*_G$ for the weight lattice, which we can consider as a subset of $\mathfrak t^*_G$. The Weyl group $W_G$ acts on $\mathfrak t_G^*$ by reflections through the hyperplanes orthogonal to the roots. Let us choose a set of positive roots $R_{G,+} \subseteq \Lambda^*_G$. This determines a positive Weyl chamber $\mathfrak t^*_{G,+}$, as well as a basis of fundamental weights $\{\omega^G_1,\ldots,\omega^G_{r_G}\}$, where $r_G = \dim T_G$ is the rank of the Lie group, and the Weyl vector $\rho = \frac 1 2 \sum_{\alpha \in R_{G,+}} \alpha$. The set of dominant weights $\Lambda^*_{G,+}$ is by definition the intersection of the weight lattice and the positive Weyl chamber.

The fundamental theorem of the representation theory of compact connected Lie groups is the fact that the irreducible (complex) representations of $G$ can be labeled by their \emph{highest weight} $\lambda \in \Lambda^*_{G,+}$ \cite{knapp02}; for every element $\lambda \in \Lambda^*_{G,+}$ there exists a unique irreducible representation $V_{G,\lambda}$ with this highest weight. Given an arbitrary finite-dimensional (complex) $G$-representation $V$, we can always decompose it into irreducible sub-representations
$
  V \cong \bigoplus_{\lambda \in \Lambda^*_{G,+}} m_{G,V}(\lambda) \, V_{G,\lambda}
$.
We shall call the function $m_{G,V}$ thus defined the \emph{highest weight multiplicity function}.

If we restrict the representation to the maximal torus, we can similarly decompose into irreducible representations. Since $T_G$ is a compact Abelian group, we can always jointly diagonalize its action, and it follows that the irreducible representations are one-dimensional. The joint eigenvalues can be encoded as a weight $\beta \in \Lambda^*_G$, and we will denote the corresponding irreducible representation of $T_G$ by $\CC_\beta$. The decomposition
$
  V \cong \bigoplus_{\beta \in \Lambda^*_G} m_{T_G,V}(\beta) \, \CC_\beta
$
then defines the \emph{weight multiplicity function} $m_{T_G,V}$.
We also set $[k] = \{ 1,\ldots,k \}$, and write $f \sim g$ for the asymptotic equivalence $\lim_{k \rightarrow \infty} f(k)/g(k) = 1$.


An equivalent way of encoding weight multiplicities is in terms of the (formal) \emph{character},
\begin{equation*}
  \ch V = \sum_\beta m_{T_G,V}(\beta) \, e^\beta,
\end{equation*}
which can be understood as the generating function of $m_{T_G,V}$. Formally, $\ch V$ is an element of the group ring $\ZZ[\Lambda^*_G]$, which consists of (finite) linear combinations of basis elements $e^\beta$ subject to the relation $e^\beta e^{\beta'} = e^{\beta + \beta'}$.
The character of an irreducible representation $V_{G,\lambda}$ is given by the \emph{Weyl character formula} \cite[p.~319]{knapp02}, 
\begin{equation}
  \label{weyl character formula}
  \ch V_{G,\lambda} =
  \frac {\sum_{w \in W_G} \det(w) \, e^{w(\lambda + \rho)}} {e^\rho \prod_{\alpha \in R_{G,+}} \left( 1 - e^{-\alpha} \right)}.
\end{equation}
Observe that we have
\begin{equation}
  \label{derivation kostant partition function}
  \begin{aligned}
  \frac 1 {\prod_{\alpha \in R_{G,+}} \left( 1 - e^{-\alpha} \right)}
  &= \prod_{\mathclap{\alpha \in R_{G,+}}} \left( 1 + e^{-\alpha} + e^{-2\alpha} + \ldots \right) \\
  &= \sum_{\mathclap{\beta \in \Lambda^*_G}} \phi_{R_{G,+}}(\beta) e^{-\beta},
  \end{aligned}
\end{equation}
where $\phi_{R_{G,+}}$ is the \emph{Kostant partition function} given by the formula
\begin{equation}
  \label{kostant partition function}
  \phi_{R_{G,+}}(\beta) = \# \{ (x_j) \in \ZZ^{\abs{R_{G,+}}}_{\geq 0} : \sum_j x_j \alpha_j = \beta \}.
\end{equation}
That is, $\phi_{R_{G,+}}$ counts the number of ways that a weight can be written as a sum of positive roots (this number is always finite since the positive roots span a proper cone).
It follows directly from \eqref{weyl character formula} and \eqref{derivation kostant partition function} and that
\begin{align*}
  \ch V_{G,\lambda}
  = &\sum_{w \in W_G} \det(w) \sum_{\beta \in \Lambda^*_G} \phi_{R_{G,+}}(\beta) e^{w(\lambda + \rho) - \rho - \beta} \\
  = &\sum_{\beta \in \Lambda^*_G} \sum_{w \in W_G} \det(w) \, \phi_{R_{G,+}}(w(\lambda + \rho) - \rho - \beta) e^\beta.
\end{align*}
In other words, the multiplicity of a weight $\beta$ in an irreducible representation $V_{G,\lambda}$ is given by the well-known \emph{Kostant multiplicity formula} \cite{kostant59},
\begin{equation}
  \label{kostant multiplicity formula}
  m_{T_G,V_{G,\lambda}}(\beta) = \sum_{w \in W_G} \det(w) \, \phi_{R_{G,+}}(w(\lambda + \rho) - \rho - \beta).
\end{equation}

For any fixed group $G$, the Kostant partition function can be evaluated efficiently by using Barvinok's algorithm \cite{barvinok94}, since it amounts to counting points in a convex polytope in an ambient space of fixed dimension. Therefore, weight multiplicities for fixed groups $G$ can be computed efficiently. This idea has been implemented by Cochet \cite{cochet05} to compute weight multiplicities for the classical Lie algebras (using the method presented in \cite{welledabaldonibeckcochetetal06} instead of Barvinok's algorithm). We remark that the problem of computing weight multiplicities is of course the special case of \autoref{main problem} where $H$ is the maximal torus $T_G \subseteq G$.

\subsection*{Weight Multiplicities as a Single Partition Function}

If $G$ is semisimple, we can find $s, t \in \ZZ_{\geq 0}$ and group homomorphisms $A \colon \ZZ^s \rightarrow \ZZ^t$ and $B \colon \Lambda^*_G \oplus \Lambda^*_G \rightarrow \ZZ^t$ such that
\begin{equation}
  \label{bliem multiplicity formula}
  m_{T_G,V_{G,\lambda}}(\beta) = \phi_A \left( B \matrix{\lambda \\ \beta} \right)
  \>\>
  (\forall \lambda \in \Lambda^*_{G,+}, \beta \in \Lambda^*_G),
\end{equation}
where $\phi_A$ is the \emph{vector partition function} defined by
\begin{equation}
  \phi_A(y) = \# \{ x \in \ZZ^s_{\geq 0} : A x = y \}.
\end{equation}
Note that this improves over the Kostant multiplicity formula \eqref{kostant multiplicity formula}, where weight multiplicities are expressed as an alternating sum over vector partition functions. In particular, \eqref{bliem multiplicity formula} is an evidently positive formula.
It has been established by Billey, Guillemin, and Rassart for the Lie algebra $\su(d)$ \cite{billeyguilleminrassart04}, and was later extended to the general case by Bliem \cite{bliem08} by considering Littelmann patterns \cite{littelmann98} instead of Gelfand--Tsetlin patterns \cite{gelfandtsetlin88}.


The assumption of semisimplicity for \eqref{bliem multiplicity formula} is not a restriction. Indeed, if $G$ is a general compact connected Lie group then its Lie algebra can always decomposed as
\begin{equation}
  \label{compact decomposition}
  \mathfrak g = [\mathfrak g,\mathfrak g] \oplus \mathfrak z,
\end{equation}
where the commutator $[\mathfrak g, \mathfrak g]$ is the Lie algebra of a compact connected semisimple Lie group $G_{\sesi}$, and where $\mathfrak z$ the Lie algebra of the center $Z(G)$ of $G$ \cite[Corollary 4.25]{knapp02}. Let us choose a maximal torus $T_{G_{\sesi}}$ of $G_{\sesi}$ that is contained in $T_G$.
Consider now an irreducible $G$-representation $V_{G,\lambda}$ with highest weight $\lambda$. By Schur's lemma, each element in $Z(G)$ acts by a scalar. Therefore, all weights $\beta$ that appear in the weight-space decomposition have the same restriction to $\mathfrak z$. It follows that
\begin{equation}
  \label{semisimple restriction}
    m_{T_G,V_{G,\lambda}}(\beta) = \begin{cases}
    m_{T_{G_{\sesi}},V_{G_{\sesi},\lambda_{\sesi}}}(\beta_{\sesi}) & \text{if $\lambda_z = \beta_z$},\\
    0 & \text{otherwise},
  \end{cases}
\end{equation}
where we write $\mu_{\sesi}$ and $\mu_z$ for the restriction of a weight $\mu$ to the Cartan subalgebra of $[\mathfrak g,\mathfrak g]$ and to $\mathfrak z$, respectively.
These multiplicities can therefore be evaluated by using \eqref{bliem multiplicity formula}.

\section{The Finite Difference Formula}
\label{section:finite difference formula}

Let $V$ be an arbitrary finite-dimensional representation of the compact, connected Lie group $G$.
Clearly, we can compute the weight multiplicity function $m_{T_G,V}$ from the highest weight multiplicity function $m_{G,V}$ by using any of the classical formulas \eqref{weyl character formula} and \eqref{kostant multiplicity formula}, or by evaluating the vector partition function \eqref{bliem multiplicity formula} described in \autoref{section:preliminaries}. By ``inverting'' the Weyl character formula, the converse can also be achieved:

\begin{prp}
  \label{steinberg lemma}
  The highest weight and weight multiplicity function of a finite-dimensional $G$-representation $V$ are related by
  \begin{equation*}
    m_{G,V} = \restrict{\left(\prod_{\alpha \in R_{G,+}} - D_\alpha \right) m_{T_G,V}}{\Lambda^*_{G,+}},
  \end{equation*}
  where $(D_\alpha m)(\lambda) = m(\lambda + \alpha) - m(\lambda)$ is the finite-difference operator in direction $\alpha$.
  Note that any two of the operators $D_\alpha$ commute, so that their product is independent of the order of multiplication.
\end{prp}
\begin{IEEEproof}
  By linearity, it suffices to establish the lemma for a single irreducible representation $V = V_{G,\lambda}$ of highest weight $\lambda$.
  The Weyl character formula \eqref{weyl character formula} can be rewritten in the form
  \begin{equation}
    \label{weyl for steinberg}
    \prod_{\alpha > 0} \left( 1 - e^{-\alpha} \right) \ch V_{G,\lambda} =
    \sum_{w \in W_G} \det(w) \, e^{w(\lambda + \rho) - \rho}.
  \end{equation}
  If we identify elements in $\ZZ[\Lambda^*_G]$ with functions on the weight lattice, applying finite-difference operators $D_\alpha$ corresponds to multiplication by $\left(e^{-\alpha} - 1 \right)$. Therefore, the left-hand side of \eqref{weyl for steinberg} is identified with $\left( \prod_{\alpha \in R_{G,+}} - D_\alpha \right) m_{T_G,V_{G,\lambda}}$.

  Now consider the right-hand side of \eqref{weyl for steinberg}. Since $\lambda + \rho$ is a strictly dominant weight, it is sent by any Weyl group element $w \neq 1$ to the interior of another Weyl chamber. That is, there exists a positive root $\alpha \in R_{G,+}$ such that $\langle \alpha, w(\lambda + \rho) \rangle < 0$. In particular, $w(\lambda + \rho) - \rho$ is never dominant unless $w = 1$. It follows that the restriction of $\left( \prod_{\alpha \in R_{G,+}} - D_\alpha \right) m_{T_G,V_{G,\lambda}}$ to $\Lambda^*_{G,+}$ is equal to the indicator function of $\{\lambda\}$, i.e., equal to the highest weight multiplicity function of $V_{G,\lambda}$.
\end{IEEEproof}

The idea of using \eqref{weyl character formula} for determining multiplicities of irreducible representations goes back at least to Steinberg \cite{steinberg61}, who proved a formula for the multiplicity $c_{\lambda,\mu}^\nu$ of an irreducible representation $V_{G,\nu}$ in the tensor product $V_{G,\lambda} \otimes V_{G,\mu}$. These multiplicities $c_{\lambda,\mu}^\nu$ are called the \emph{Littlewood--Richardson coefficients} for $G$. Steinberg's formula involves an alternating sum over the Kostant partition function \eqref{kostant multiplicity formula}; it can be evaluated efficiently as described by Cochet \cite{cochet05}. De Loera and McAllister give another method for computing Littlewood--Richardson coefficients \cite{deloeramcallister06}, which applies Barvinok's algorithm to results by Berenstein and Zelevinsky \cite{berensteinzelevinsky01}.
Since the tensor products of irreducible $G$-representations are just the irreducible representations of $G \times G$, the problem of computing Littlewood--Richardson coefficients is again a special case of \autoref{main problem}.
The following consequence of the proof of \autoref{steinberg lemma} will be convenient in the sequel:

\begin{cor}
  \label{steinberg corollary}
  Write $\prod_{\alpha \in R_{G,+}} \left( 1 - e^{-\alpha} \right) = \sum_{\gamma \in \Gamma_G} c_\gamma e^{-\gamma}$ with $\Gamma_G \subseteq \Lambda^*_G$ finite and all $c_\gamma \neq 0$. Then,
  \begin{equation*}
    m_{G,V}(\lambda) = \sum_{\gamma \in \Gamma_G} c_\gamma \, m_{T_G,V}(\lambda + \gamma).
  \end{equation*}
\end{cor}

In particular, it is evident from \autoref{steinberg corollary} that, for any fixed group $G$, the multiplicity of an irreducible representation in some representation $V$ can be computed efficiently from the weight multiplicities of $V$ by computing a finite linear combination.

\section{Multiplicities for the Subgroup Restriction Problem}
\label{section:finite difference formula for subgroup restrictions}

Every $G$-representation $V$ can be considered as (``restricts to'') a representation of $H$ by setting
\begin{equation}
  \label{restriction}
  h \cdot v := f(h) \cdot v \qquad (\forall h \in H),
\end{equation}
and the subgroup restriction problem for $f$, as defined in \autoref{main problem}, amounts to determining the multiplicity $m^\lambda_\mu$ of a given irreducible representation of $H$ in the restriction of a given irreducible representation of $G$.
In this section we will derive a formula for these multiplicities (\autoref{main theorem}), which will be the main ingredient of the algorithm presented in \autoref{section:algorithm} below.
It will also follow from this formula that the $m^\lambda_\mu$ are given by a piecewise quasi-polynomial function%
\footnote{In the context of this paper, a quasi-polynomial function is a polynomial function with periodic coefficients; see p.~\pageref{pageref:quasi-poly} for the precise definition. It should not to be confused with the notion of quasi-polynomial time complexity.}
in $\lambda$ and $\mu$ (\autoref{abstract cor}).

Let us choose the maximal torus $T_H \subseteq H$ in such a way that $f(T_H) \subseteq T_G$, and denote the corresponding Cartan subalgebra by $\mathfrak t_H$. Of course, this implies that the induced Lie algebra homomorphism $\Lie(f)$ sends the Cartan subalgebra of $H$ in the one of $G$. Since $f$ is a group homomorphism, $\Lie(f)$ restricts to a homomorphism between the integral lattices, $F \colon \Lambda_H \rightarrow \Lambda_G, ~ X \mapsto \Lie(f) X$.
The dual map between the weight lattices is given by
\begin{equation}
  \label{definition dual map}
  F^* \colon \Lambda_G^* \rightarrow \Lambda_H^*, \quad
  \beta \mapsto \beta \circ F = \restrict{\beta \circ \Lie(f)}{\Lambda_H}.
\end{equation}

The following is well-known and easily follows from the definitions:

\begin{lem}
  \label{weight restriction}
  Let $V$ be a representation of $G$ and $v \in V$ a weight vector of weight $\beta \in \Lambda^*_G$.
  If we restrict the action to $H$ via \eqref{restriction} then $v$ is a weight vector of weight $F^*(\beta) \in \Lambda^*_H$.
\end{lem}

Let us also fix systems of positive roots $R_{H,+}$ for $H$. This in turn determines the set of dominant weights $\Lambda^*_{H,+}$ as well as a basis of fundamental weights $(\omega^H_j)$ as described in \autoref{section:preliminaries}. Let us also set $r_H = \dim T_H$.

Our strategy for solving the subgroup restriction problem for $f$ then is the following:
Given an irreducible representation $V_{G,\lambda}$ of $G$, we can determine its weight multiplicities with respect to the maximal torus $T_G$ by using any of the formulas presented in \autoref{section:preliminaries}. We then obtain weight multiplicities for $T_H$ by restricting according to \autoref{weight restriction}. Finally, we reconstruct the multiplicity of an irreducible representation $V_{H,\mu}$ by using the finite-difference formula (\autoref{steinberg lemma}/\autoref{steinberg corollary}).
If this procedure was translated directly into an algorithm, the runtime would be polynomial in the coefficients of $\lambda$ (with respect to the basis of fundamental weights), i.e., exponential in their bitlength, since the number of weights is of the order of the dimension of the irreducible representation $V_{G,\lambda}$, which according to the \emph{Weyl dimension formula} is given by the polynomial $\prod_{\alpha \in R_{G,+}} {\langle \alpha, \lambda + \rho \rangle} / {\langle \alpha, \rho \rangle}$ (cf.\ the formula by Straumann \cite{straumann65}).
We will now show that it is possible to combine the weight multiplicity formula \eqref{bliem multiplicity formula} with the restriction map $F^*$ in a way that will later give rise to an algorithm that runs in polynomial time in the bitlength of the input:

\begin{thm}
  \label{main theorem}
  Let $f \colon H \rightarrow G$ be a homomorphism of compact connected Lie groups.
  Then we can find $s, s', u \in \ZZ_{\geq 0}$ and group homomorphisms $\mathcal A \colon \ZZ^{s+s'} \rightarrow \ZZ^u$ and $\mathcal B \colon \Lambda^*_G \oplus \Lambda^*_H \rightarrow \ZZ^u$ with the following property: For every irreducible representation $V_{G,\lambda}$ of $G$ and $V_{H,\mu}$ of $H$, the multiplicity $m^\lambda_\mu$ of the latter in the former is given by
  \begin{equation*}
    m^\lambda_\mu =
    \sum_{\gamma \in \Gamma_H} c_\gamma \, \# \{ x \in \ZZ^s_{\geq 0} \oplus \ZZ^{s'} : \mathcal A x = \mathcal B \Matrix{\lambda \\ \mu + \gamma} \},
  \end{equation*}
  where the (finite) set $\Gamma_H$ and the coefficients $(c_\gamma)$ are defined by $\prod_{\alpha \in R_{H,+}} \left( 1 - e^{-\alpha} \right) = \sum_{\gamma \in \Gamma_H} c_\gamma e^{-\gamma}$ and $c_\gamma \neq 0$.
  In fact, we can choose $s = O(r_G^2)$, $s' \leq r_G$ and $u = O(r_G^2) + r_H$.
\end{thm}
\begin{IEEEproof}
  By definition and \autoref{steinberg corollary}, we have
  $m^\lambda_\mu = m_{H,V_{G,\lambda}}(\mu) = \sum_{\gamma \in \Gamma_H} c_\gamma \, m_{T_H,V_{G,\lambda}}(\mu + \gamma)$.
  In view of \autoref{weight restriction}, the multiplicity of a $T_H$-weight $\delta \in \Lambda^*_H$ in the irreducible $G$-representation $V_{G,\lambda}$ is given by
  \begin{equation*}
    m_{T_H,V_{G,\lambda}}(\delta) ~=~
    \sum_{\mathclap{\substack{\beta \in \Lambda^*_G\\ F^*(\beta) = \delta}}} m_{T_G,V_{G,\lambda}}(\beta).
  \end{equation*}
  As in \eqref{compact decomposition}, let us now decompose the Lie-algebra $\mathfrak g = [\mathfrak g, \mathfrak g] \oplus \mathfrak z$.
  Denote the Lie group corresponding to $[\mathfrak g,\mathfrak g]$ by $G_{\sesi}$ and choose a maximal torus $T_{G_{\sesi}}$ which is contained in $T$.
  Using \eqref{semisimple restriction},
  \begin{align*}
    \sum_{\mathclap{\substack{\beta \in \Lambda^*_G\\ F^*(\beta) = \delta}}} m_{T_G,V_{G,\lambda}}(\beta)
    ~~=~~ \sum_{\mathclap{\substack{\beta_{\sesi} \in \Lambda^*_{G_{\sesi}}\\ C_{\sesi} \beta_{\sesi} + C_z \lambda_z = \delta}}} m_{T_{G_{\sesi}},V_{G_{\sesi},\lambda_{\sesi}}}(\beta_{\sesi}),
  \end{align*}
  where we have decomposed $F^*$ as a sum of two homomorphisms $C_{\sesi} \colon \Lambda^*_{G_{\sesi}} \rightarrow \Lambda^*_H$ and $C_z \colon \Lambda^*_{Z(G)} \rightarrow \Lambda^*_H$.

  Let us now choose group homomorphisms $A \colon \ZZ^s \rightarrow \ZZ^t$ and $B = B_1 \oplus B_2 \colon \Lambda^*_{G_{\sesi}} \oplus \Lambda^*_{G_{\sesi}} \rightarrow \ZZ^t$ such that \eqref{bliem multiplicity formula} holds for the weight multiplicities for $G_{\sesi}$.
  For this, $s$ and $t$ can be taken of order $O(r_G^2)$ \cite[Proposition 19]{bliem08}.
  Then,
  \begin{equation}
  \begin{aligned}
  \label{constructive}
    &\sum_{\mathclap{\substack{\beta_{\sesi} \in \Lambda^*_{G_{\sesi}}\\ C_{\sesi} \beta_{\sesi} + C_z \lambda_z = \delta}}} m_{T_{G_{\sesi}},V_{G_{\sesi},\lambda_{\sesi}}}(\beta_{\sesi}) \\
    =~&\sum_{\mathclap{\substack{\beta_{\sesi} \in \Lambda^*_{G_{\sesi}}\\ C_{\sesi} \beta_{\sesi} + C_z \lambda_z = \delta}}}
      \# \{ x \in \ZZ^s_{\geq 0} : A x = B \matrix{\lambda_{\sesi} \\ \beta_{\sesi}} \} \\
    =~&\# \{ (x,\beta_{\sesi})
      :
      \matrix{
        A & - B_2\\
        0 & C_{\sesi}
      }
      \matrix{x \\ \beta_{\sesi}} =
      \matrix{B_1 \lambda_{\sesi} \\ - C_z \lambda_z + \delta} \}\\
    =~&\# \{ (x,\beta_{\sesi})
      :
      \underbrace{\matrix{
        A & - B_2\\
        0 & C_{\sesi}
      }}_{=: \mathcal A}
      \matrix{x \\ \beta_{\sesi}} =
      \underbrace{\matrix{
        B_1 & 0 & 0\\
        0 & -C_z & \Id
      }}_{=: \mathcal B}
      \matrix{\lambda_{\sesi} \\ \lambda_z \\ \delta} \}.
  \end{aligned}
  \end{equation}
  After choosing a basis of the lattice $\Lambda^*_{G_{\sesi}}$ we arrive at the asserted formula (with $s' = \dim T_{G_{\sesi}}$ and $u = t + r_H$).
\end{IEEEproof}

We stress that the proof of \autoref{main theorem} is constructive: The maps $\mathcal A$ and $\mathcal B$, whose existence is asserted by the theorem, are defined in \eqref{constructive} in terms of $A$ and $B$, whose construction is described explicitly in \cite[Proof of Theorem 2.1]{billeyguilleminrassart04} (for the case of $\mathfrak g = \su(d)$) and in \cite[\S 4]{bliem08} (for the general case). See \autoref{section:kronecker} for an illustration in the context of the Kronecker coefficients.

If one uses the Kostant multiplicity formula \eqref{kostant multiplicity formula} instead of \eqref{bliem multiplicity formula} in the proof of \autoref{main theorem} then one arrives at a similar formula for the multiplicities $m^\lambda_\mu$ involving an additional alternating sum over the Weyl group of $G$. After completion of this work, we have learned of \cite[Lemma 3.1]{heckman82} which is derived in this spirit.

\subsection*{Piecewise Quasi-Polynomiality}

Let us use the fundamental weight bases fixed above to identify $\Lambda^*_G \cong \ZZ^{r_G}$ and $\Lambda^*_H \cong \ZZ^{r_H}$. The group homomorphisms $\mathcal A$ and $\mathcal B$ correspond to matrices with integer entries, which we shall denote by the same symbols.
Observe that the formula in \autoref{main theorem} in essence amounts to counting the number
$n(y) := \# \left( \Delta_{\mathcal A, \mathcal B}(y) \cap \ZZ^{s+s'} \right)$
of integral points in certain rational convex polytopes of the form
\begin{equation}
  \label{weight restriction polytope}
  \Delta_{\mathcal A,\mathcal B}(y) := \{ x \in \RR^{s+s'} : x_1, \ldots, x_s \geq 0, \mathcal A x = \mathcal B y \},
\end{equation}
parametrized by $y \in \ZZ^{r_G+r_H}$. Explicitly,
\begin{equation}
  \label{concrete alternating sum}
  m^\lambda_\mu = \sum_{\gamma \in \Gamma_H} c_\gamma n(\lambda, \mu + \gamma).
\end{equation}
It is well-known that $n(y)$ is a \emph{piecewise quasi-polynomial} function in $y$ \cite{claussloechner98}.
\label{pageref:quasi-poly}
That is, there exists a decomposition of $\ZZ^{r_G+r_H}$ into polyhedral chambers such that on each chamber $C$ the function $n(y)$ is given by a single quasi-polynomial, i.e., there exists a sublattice $L \subseteq \ZZ^{r_G+r_H}$ of finite index and polynomials $(p_z)$ with rational coefficients, labeled by the finitely many points $z \in \ZZ^{r_G+r_H} / L$, such that
$n(y) = p_{[y]}(y)$
for all $y \in \ZZ^{r_G+r_H}$ (cf.\ \cite[\S 2.2]{verdoolaegeseghirbeylsetal07}). We record the following immediate consequence:

\begin{cor}
  \label{abstract cor}
  For any fixed group homomorphism $f \colon H \rightarrow G$, the multiplicities $m^\lambda_\mu$ are given by a piecewise quasi-polynomial function in $\lambda$ and $\mu$.
\end{cor}

In particular, this implies that the \emph{stretching function} $k \mapsto m^{k\lambda}_{k\mu}$ is a quasi-polynomial function for large $k$. This is in fact true for all $k$, as has been observed in \cite{mulmuley07} (cf.~\cite{meinrenkensjamaar99} for more general quasi-polynomiality results on convex cones, and also \cite{baldonivergne10} for further discussion).

\section{Polynomial-Time Algorithm for the Subgroup Restriction Problem}
\label{section:algorithm}

In this section we will formulate our algorithm for the subgroup restriction problem, \autoref{main problem}.
Recall that, by \eqref{concrete alternating sum}, the computation of the multiplicities $m^\lambda_\mu$ effectively reduces to counting the number of integral points in certain rational convex polytopes of the form \eqref{weight restriction polytope}.
We shall suppose that the highest weights $\lambda$ and $\mu$, which are the \emph{input} to our algorithm, are given in terms of their coordinates with respect to the fundamental weight bases fixed in \autoref{section:finite difference formula for subgroup restrictions}. Clearly, for each of the finitely many $\gamma \in \Gamma_H$, the description of the polytope $\Delta_{\mathcal A,\mathcal B}(\lambda,\mu+\gamma)$ (say, in terms of linear inequalities) is of polynomial size in the bitlength of the input. It follows that \emph{Barvinok's algorithm} can be used to compute the number of integral points in each of these polytopes in polynomial time \cite{barvinok94} (see also \cite{dyerkannan97,barvinokpommersheim99}).
This gives rise to the following polynomial-time algorithm for \autoref{main problem}, thereby establishing \autoref{A}:

\begin{alg}
  \label{main algorithm}
  Let $f \colon H \rightarrow G$ be a homomorphism of compact connected Lie groups.
  Given as input two highest weights $\lambda \in \Lambda^*_G \cong \ZZ^{r_G}$ and $\mu \in \Lambda^*_H \cong \ZZ^{r_H}$, encoded as bitstrings containing their coordinates with respect to the fundamental weight bases fixed above, the following algorithm computes the multiplicity $m^\lambda_\mu$ in polynomial time in the bitlength of the input:
  \begin{algorithmic}
    \State $m \gets 0$
    \ForAll{$\gamma \in \Gamma_H$}
      \State $n \gets \# \left( \Delta_{\mathcal A,\mathcal B}(\lambda, \mu + \gamma) \cap \ZZ^{s+s'} \right)$ as computed by Barvinok's algorithm (see discussion above)
      \State $m \gets m + c_\gamma n$
    \EndFor
    \State \textbf{return} $m$
  \end{algorithmic}
  Here, $\Delta_{\mathcal A,\mathcal B}(y)$ denotes the rational convex polytope defined in \eqref{weight restriction polytope}, and the finite index set $\Gamma_H \subseteq \Lambda^*_H$ as well as the coefficients $(c_\gamma)$ are defined in the statement of \autoref{main theorem}.
\end{alg}

There are at least two software packages which have implemented Barvinok's algorithm, namely \textsc{LattE} \cite{deloeradutrakoppeetal11} and \textsc{barvinok} \cite{verdoolaegeseghirbeylsetal07,verdoolaegebruynooghe08}. In \autoref{section:introduction} we have reported on the performance of our implementation of \autoref{main algorithm} for computing Kronecker coefficients using the latter package.

\begin{rem}
  The existence of a polynomial-time algorithm for \autoref{main problem} in fact already follows abstractly from \autoref{abstract cor}, since in order to compute $m^\lambda_\mu$ we merely have to evaluate a \emph{fixed} piecewise quasi-polynomial function. This piecewise quasi-polynomial can be computed algorithmically by using a variant of Barvinok's algorithm which is also implemented in the \textsc{barvinok} package; see \cite[Proposition 2]{verdoolaegeseghirbeylsetal07} and also \cite[(5.3.1)]{barvinokpommersheim99}.
\end{rem}


\section{Kronecker Coefficients}
\label{section:kronecker}

As explained in the introduction, the Kronecker coefficients play an important role in geometric complexity theory and quantum information theory. In this section, we will describe precisely how they can be computed using our methods.

Let us recall the language of Young diagrams which is commonly used in this context \cite{fulton97}. A \emph{Young diagram} with $r$ rows and $k$ boxes is given by an ordered list of integers $\lambda_1 \geq \ldots \geq \lambda_r > 0$ with $\sum_i \lambda_i = k$.
It can be visualized as an arrangement of $k$ boxes in $r$ rows with $\lambda_j$ boxes in the $j$-th row.
We set $\lambda_j = 0$ for all $j > r$.
We will now consider the \emph{unitary group} $\U(d)$, which consists of the unitary $d \times d$-matrices. Let us fix a system of positive roots and denote the corresponding basis of fundamental weights by $(\omega_j)$.
To each Young diagram $\lambda$ with at most $d$ rows we associate the irreducible representation of $\U(d)$ with highest weight equal to $\sum_{j=1}^d \left( \lambda_j - \lambda_{j+1} \right) \omega_j$.
Every polynomial irreducible representation of $\U(d)$ arises in this way.
By a slight abuse of notation, we identify Young diagrams with the corresponding highest weights.
More generally, we can associate to every integer vector $\beta \in \ZZ^d$ the weight $\sum_{j=1}^d \left( \beta_j - \beta_{j+1} \right) \omega_j$, where we set $\beta_{d+1} = 0$.
This defines a bijection between $\ZZ^d$ and the weight lattice $\Lambda^*_{\U(d)}$ of $\U(d)$. In particular, the positive roots fixed above correspond to the integer vectors of the form $(\ldots,0,1,0,\ldots,0,-1,0,\ldots)$.

\bigskip

The \emph{Kronecker coefficient} $g_{\lambda,\mu,\nu}$ associated with triples of Young diagrams $\lambda$, $\mu$ and $\nu$ with $k$ boxes each and at most $a$, $b$ and $c$ rows, respectively, can then be defined in terms of the following subgroup restriction problem of compact, connected Lie groups:
Let $H = \U(a) \times \U(b) \times \U(c)$ and $G = \U(abc)$ and consider the homomorphism $f \colon H \rightarrow G$ given by sending a triple of unitaries $(U,V,W)$ to their tensor product $U \otimes V \otimes W$. The Kronecker coefficient $g_{\lambda,\mu,\nu}$ is then given by the multiplicity of the irreducible $H$-representation $V_{H,(\lambda,\mu,\nu)} = V_{\U(a),\lambda} \otimes V_{\U(b),\mu} \otimes V_{\U(c),\nu}$ in the restriction of the symmetric power $\Sym^k(\CC^{abc})$, which is the irreducible $G$-representation labeled by the Young diagram $(k)$ consisting of a single row with $k$ boxes. That is,
\begin{equation}
  \label{kronecker definition}
  g_{\lambda,\mu,\nu} = m^{(k)}_{\lambda,\mu,\nu}(f)
\end{equation}
This definition in fact does not depend on the concrete values chosen for $a$, $b$ and $c$, as can be seen by rephrasing it in terms of the representation theory of the symmetric group $S_k$ \cite[\S 8]{burgisserlandsbergmaniveletal11} (but of course $a$, $b$ and $c$ have to be chosen at least as large as the number of rows of the Young diagrams). Moreover, it is evident that the Kronecker coefficients are symmetric in the variables $\lambda$, $\mu$, and $\nu$.

It follows that, for any fixed choice of $a$, $b$ and $c$, \autoref{main algorithm} can be used to compute the Kronecker coefficient \eqref{kronecker definition} given Young diagrams with at most $a$, $b$ and $c$ rows, respectively, in polynomial time in the input size, or equivalently in time $O(\poly(\log k))$, where $k$ is the number of boxes of the Young diagrams. This establishes \autoref{B}.
Let us again stress that the problem of computing Kronecker coefficients is known to be $\SharpP$-hard in general \cite{burgisserikenmeyer08}; hence we do not expect that there exists a polynomial-time algorithm without any assumption on the number of rows of the Young diagrams.

When computing Kronecker coefficients using the above method, we are only interested in the representation $V_{\U(abc),(k)} = \Sym^k(\CC^{abc})$, not in arbitrary irreducible representations of $\U(abc)$. By specializing the construction described in \autoref{main theorem} to this one-parameter family of representations, we obtain the following result:

\begin{prp}
  \label{optimized kronecker}
  The multiplicity of a weight $\delta = (\delta^{A},\delta^B,\delta^C) \in \ZZ^a \oplus \ZZ^b \oplus \ZZ^c \cong \Lambda^*_H$ (we use the identifications fixed at the beginning of \autoref{section:kronecker}) in the irreducible $G$-representation $\Sym^k(\CC^{a b c})$ is equal to the number of integral points in the rational convex polytope
  \begin{equation*}
  \begin{aligned}
    &\Delta(k, \delta)
    = \Big\{
      (x_{l,m,n}) \in \RR^{a b c}_{\geq 0} \;:\; \sum_{l,m,n} x_{l,m,n} = k, \\
    &\quad \sum_{m,n} x_{l,m,n} = \delta^{A}_l,
      \sum_{l,n} x_{l,m,n} = \delta^{B}_m,
      \sum_{l,m} x_{l,m,n} = \delta^{C}_n
    \Big\}.
  \end{aligned}
  \end{equation*}
  It follows that the Kronecker coefficient for Young diagrams $\lambda, \mu, \nu$ with $k$ boxes and at most $a$, $b$ and $c$ rows, respectively, is given by the formula
  \begin{equation*}
    g_{\lambda,\mu,\nu} = \sum_{\gamma \in \Gamma_H} c_\gamma \, \# \left( \Delta(k, (\lambda,\mu,\nu)+\gamma) \cap \ZZ^{abc} \right),
  \end{equation*}
  where $\Gamma_H$ and $(c_\gamma)$ are defined as in the statement of \autoref{steinberg corollary}.
\end{prp}
\begin{IEEEproof}
  It is well-known that the weight spaces for the action of $\U(d)$ on $\Sym^k(\CC^d)$ are all one-dimensional and that the set of weights corresponds to the integer vectors in the standard simplex rescaled by $k$ \cite{fulton97}.
  In our case, $d = a b c$, so that the weights are just the integral points of the polytope
  \begin{equation*}
    \Big\{ x = (x_{l,m,n})_{l \in [a], m \in [b], n \in [c]} \in \RR^{a b c}_{\geq 0} \;:\; \sum_{l,m,n} x_{l,m,n} = k \Big\}.
  \end{equation*}
  Moreover, the dual map $F^* \colon \Lambda^*_{\U(abc)} \rightarrow \Lambda^*_{\U(a) \times \U(b) \times \U(c)}$ as defined in \eqref{definition dual map} is given by
  \begin{equation*}
    \begin{cases}
    \ZZ^{abc} &\rightarrow \ZZ^a \oplus \ZZ^b \oplus \ZZ^c \\
    (x_{l,m,n}) &\mapsto \Big( \sum_{m,n} x_{l,m,n}, \sum_{l,n} x_{l,m,n}, \sum_{l,m} x_{l,m,n} \Big).
    \end{cases}
  \end{equation*}
  We conclude that the multiplicity of a weight $\delta = (\delta^A, \delta^B, \delta^C)$ for $\U(a) \times \U(b) \times \U(c)$ is given by the number of integral points in the polytope $\Delta(k,\delta)$ described above.
\end{IEEEproof}

Just as for our main algorithm, \autoref{optimized kronecker} gives rise to a polynomial-time algorithm for computing Kronecker coefficients with a bounded number of rows.
This second algorithm runs faster than the generic one presented earlier, since the ambient space $\RR^{abc}$ has a smaller dimension than what we would get from the construction described in the proof of \autoref{main theorem}.
We remark that the time complexity for unbounded $a$, $b$ and $c$ can be deduced from \cite{barvinokpommersheim99}.


\section{Asymptotics}
\label{section:asymptotics}

In this section we will prove our result on the generic order of growth of multiplicities in the coordinate ring of a projective variety (\autoref{C}).

We will work in the following general setup: Let $V$ be a finite-dimensional rational representation of $H$, and suppose that $X$ is an $H$-stable closed subvariety of the associated projective space $\PP(V)$. The homogeneous coordinate ring $\CC[X]$ is graded, and we can decompose each part into its irreducible components,
\begin{equation}
  \label{alggeo setup}
  \CC[X]
  = \bigoplus_{k=0}^\infty \CC[X]_k
  = \bigoplus_{k=0}^\infty \bigoplus_{\mu} m_{H,X,k}(\mu) \, V_{H,\mu}^*,
\end{equation}
where, following the usual conventions, we have decomposed with respect to the dual representations $V_{H,\mu}^*$. The \emph{stretching function} is then by definition $k \mapsto m_{H,X,k}(k\mu)$. We stress that in contrast to \cite{mulmuley07}, where it was assumed that $X$ has at most rational singularities, we do not even require that $X$ is a normal variety \cite{hartshorne77}. This is highly relevant for geometric complexity theory, since it was recently shown in \cite{kumar10b} and \cite{burgisserikenmeyer11} that the studied varieties (the orbit closures of the determinant and permanent on the one hand, and of the matrix multiplication tensor and the unit tensor on the other hand) are in fact never normal except in trivial situations.

\begin{rem}
  The subgroup restriction problem for a rational group homomorphism $f \colon H \rightarrow G$ can be realized in the above setup:
  Indeed, for any highest weight $\lambda \in \Lambda^*_{G,+}$ consider $X = \mathcal O_{G,\lambda}$, the coadjoint orbit through $\lambda$, with the induced action of $H$. This variety can be canonically embedded into projective space as the orbit of the highest weight vector in $\PP(V_{G,\lambda})$,
  and it is a consequence of the Borel--Weil theorem that $\CC[\mathcal O_{G,\lambda}] = \bigoplus_{k=0}^\infty V_{G,k\lambda}^*$.
  By comparing with \eqref{alggeo setup} it follows that $m_{H,\mathcal O_{G,\lambda},k}(\mu) = m^{k \lambda}_{\mu}$. In particular, the above definition of the stretching function, $k \mapsto m_{H,\mathcal O_{G,\lambda},k}(k\mu)$, coincides with our previous usage, $k \mapsto m^{k\lambda}_{k\mu}$.
\end{rem}

\begin{IEEEproof}[Proof of \autoref{C}]
  By the Hilbert--Serre theorem, the function $k \mapsto \dim \CC[X]_k$ is a polynomial of degree $\dim X$ for large $k$ \cite[Theorem I.7.5]{hartshorne77}. Hence there exists a constant $A > 0$ such that
  \begin{align*}
    &A \, k^{\dim X}
    \sim \dim \CC[X]_k \\
    ~=~ &\sum_{\mathclap{\mu \in \Lambda^*_{H,+}}} m_{H,X,k}(\mu) \, \dim V_\mu
    ~=~ \sum_{\mathclap{\mu \in \Delta_X \cap \frac 1 k \Lambda^*_{H,+}}} m_{H,X,k}(k \mu) \, \dim V_{k \mu},
  \end{align*}
  where for the last equality we have used the definition of the moment polytope $\Delta_X$.
  By the Weyl dimension formula, we have
  \begin{align*}
    &\dim V_{k\mu} ~=~
    \prod_{\mathclap{\alpha \in R_{H,+}}} \frac {\langle \alpha, k \mu + \rho \rangle} {\langle \alpha, \rho \rangle} \\
    ~=~ &\left( ~~~
      \prod_{\mathclap{\substack{\alpha \in R_{H,+} \\ \alpha \not\perp \Delta_X}}} \frac {\langle \alpha, \mu \rangle} {\langle \alpha, \rho \rangle}
    \right) k^{R_X} +
    O(k^{R_X-1})
  \end{align*}
  for the representations that occur in $\CC[X]$. The coefficient $P(\mu) = \prod_{\alpha \in R_{H,+}, \alpha \not\perp \Delta_X} {\langle \alpha, \mu \rangle} / {\langle \alpha, \rho \rangle}$ is a polynomial function in $\mu$. Since $\Delta_X$ is compact, we can therefore find a constant $C > 0$ such that
  \begin{equation*}
    \dim V_{k \mu} \leq C \, k^{R_X}   \qquad (\forall k, \mu \in \Delta_X \cap \frac 1 k \Lambda^*_{H,+}).
  \end{equation*}
  It follows that
  \begin{align*}
    &\sum_{\mathclap{\mu \in \Delta_X \cap \frac 1 k \Lambda^*_{H,+}}} m_{H,X,k}(k \mu) \, \dim V_{k \mu} \\
    \leq~&C \, k^{R_X} \sum_{\mathclap{\mu \in \Delta_X \cap \frac 1 k \Lambda^*_{H,+}}} m_{H,X,k}(k \mu)
    \sim
    C \, k^{R_X + d_X} \int d\DuHe_X,
  \end{align*}
  so that $\dim X \leq R_X + d_X$.

  On the other hand, since $\DuHe_X$ is Lebesgue-absolutely continuous, the boundary of the moment polytope does not carry any measure. We can therefore find a compact set $K$ contained in the (relative) interior of the moment polytope which has positive measure with respect to $\DuHe_X$.
  Note that $P(\mu)$ is positive for all $\mu$ contained in the interior of the moment polytope (indeed, for all positive roots $\alpha$ with $\alpha \not\perp \Delta_X$ there exists $\nu \in \Delta_X$ such that $\langle \alpha, \nu \rangle > 0$; since we can always write $\mu$ as a proper convex combination of $\nu$ and some other point $\nu' \in \Delta_X$, it follows that $\langle \alpha, \mu \rangle > 0$). This implies that on the compact set $K$ we can bound $P(\mu)$ from below by a positive constant. Thus there exists a constant $D > 0$ (depending on $K$) such that
  \begin{equation*}
    \dim V_{k \mu} \geq D \, k^{R_X}   \qquad (\forall \mu \in K \cap \frac 1 k \Lambda^*_{H,+}).
  \end{equation*}
  Consequently,
  \begin{align*}
    &\sum_{\mathclap{\mu \in \Delta_X \cap \frac 1 k \Lambda^*_{H,+}}} m_{H,X,k}(k \mu) \, \dim V_{k \mu}
    \geq~\sum_{\mathclap{\mu \in K \cap \frac 1 k \Lambda^*_{H,+}}} m_{H,X,k}(k \mu) \, \dim V_{k \mu} \\
    \geq~&D \, k^{R_X} \sum_{\mathclap{\mu \in K \cap \frac 1 k \Lambda^*_{H,+}}} m_{H,X,k}(k \mu)
    \sim D \, k^{R_X + d_X} \int_K d\DuHe_X.
  \end{align*}
  We conclude that also $\dim X \geq R_X + d_X$, hence we have equality.
\end{IEEEproof}

\bigskip

Let us now elaborate on the argument presented at the end of the introduction, where we showed that Duistermaat--Heckman measures do not directly give rise to new complexity-theoretic obstructions. For this, we consider a pair of projective subvarieties $X$ and $Y$ with $\dim X < \dim Y$, as is the case for the orbit closures of relevance to GCT.
Let us assume that $\Delta_X \subseteq \Delta_Y$, so that the moment polytopes alone do not already give rise to an obstruction.
Clearly, this implies that $R_X \leq R_Y$.

\begin{lem}
  \label{polytope smaller lemma}
  Let $\Delta_X \subseteq \Delta_Y$ and $R_X < R_Y$. Then, $\dim \Delta_X < \dim \Delta_Y$.
\end{lem}
\begin{IEEEproof}
  Note that we have
  \begin{equation*}
    \dim \Delta_X = \dim \aff \Delta_X \leq \dim \aff \Delta_Y = \dim \Delta_Y,
  \end{equation*}
  with equality if and only if the two affine hulls $\aff \Delta_X \subseteq \aff \Delta_Y$ are equal.

  Now by assumption there exists a positive root $\alpha \in R_{H,+}$ that is orthogonal to all points in $\Delta_X$ (i.e., for all $p \in \Delta_X$, $\alpha \perp p$), but not to all points in $\Delta_Y$.
  It follows that $\alpha$ is also orthogonal to all points in the affine hull of $\Delta_X$, but not to all points in the affine hull of $\Delta_Y$.
  Therefore, we have $\aff \Delta_X \subsetneq \aff \Delta_Y$.
\end{IEEEproof}

\begin{lem}
  \label{different lebesgue lemma}
  Let $\dim \Delta_X < \dim \Delta_Y$. Then, $X \subseteq Y$ implies $d_X < d_Y$.
\end{lem}
\begin{IEEEproof}
  If $X \subseteq Y$ then it is immediate from \eqref{mult crit} and \eqref{definition duistermaat-heckman} that $d_X \leq d_Y$.
  Let us suppose for a moment that in fact $d_X = d_Y$. Then it follows from \eqref{mult crit} that
  \begin{equation*}
    \int_{\Delta_X} d\DuHe_X(\mu) \, g(\mu) \leq
    \int_{\Delta_Y} d\DuHe_Y(\mu') \, g(\mu')
  \end{equation*}
  for any test function $g$. In particular, this inequality would hold for $g$ the indicator function of $\Delta_X$.
  But this is clearly impossible, since $\DuHe_Y$ is absolutely continuous with respect to Lebesgue measure on $\Delta_Y$, for which $\Delta_X$ is a set of measure zero.
\end{IEEEproof}

\begin{cor}
  \label{dimension lemma}
  Let $\dim X < \dim Y$. Then, $X \subseteq Y$ implies $d_X < d_Y$.
\end{cor}
\begin{IEEEproof}
  Clearly, $X \subseteq Y$ implies that $\Delta_X \subseteq \Delta_Y$ and $R_X \leq R_Y$. If $R_X = R_Y$ then the assertion follows directly from \autoref{C}, since
  \begin{equation*}
    d_X = \dim X - R_X < \dim Y - R_Y = d_Y.
  \end{equation*}
  Otherwise, if $R_X < R_Y$, it follows from combining \autoref{polytope smaller lemma} and \autoref{different lebesgue lemma}.
\end{IEEEproof}

As described in the introduction, the upshot of the above is that we cannot directly deduce from \eqref{mult crit} a new criterion for obstructions based on the Duistermaat--Heckman measure that goes beyond what is provided by the moment polytope.

\section*{Acknowledgements}

We would like to thank Aravind Asok, Emmanuel Briand, Peter B\"urgisser, David Gross, Christian Ikenmeyer, Stavros Kousidis, Graeme Mitchison, Mercedes Rosas, Volkher Scholz, and Mich\`{e}le Vergne for helpful discussions.

This work is supported by the Swiss National Science Foundation (grant PP00P2--128455 and 200021\_138071), the German Science Foundation (grants CH 843/1--1 and CH 843/2--1), and the National Center of Competence in Research `Quantum Science and Technology'.


\bibliographystyle{IEEEtran}
\bibliography{multiplicities}

\end{document}